\documentclass[preprint,5p,times,twocolumn]{elsarticle}
\usepackage{graphicx,color,amsmath,chemformula} 
\usepackage{lineno,blkarray,multirow}
\usepackage[pagebackref=false,colorlinks,linkcolor=blue,citecolor=blue,urlcolor=blue]{hyperref}

\journal{Materials Today Communications}
\newcommand{\BX}{Cs$_2$Tl$BX_6$}
\newcommand{\BiX}{Cs$_2$TlBi$X_6$}
\newcommand{\BiCl}{Cs$_2$TlBiCl$_6$}
\newcommand{\BiBr}{Cs$_2$TlBiBr$_6$}
\newcommand{\BiI}{Cs$_2$TlBiI$_6$}
\newcommand{\InX}{Cs$_2$TlIn$X_6$}
\newcommand{\InCl}{Cs$_2$TlInCl$_6$}
\newcommand{\InBr}{Cs$_2$TlInBr$_6$}
\newcommand{\InI}{Cs$_2$TlInI$_6$}

\begin{document}

\begin{frontmatter}
\title{Theoretical investigations of electronic and optical properties of double perovskite Cs$_2$Tl$BX_6$ ($B=$ Bi, In; $X=$ Cl, Br, I) for photovoltaic application}

\author[brin]{Ardimas}
\ead{ardimas2811@gmail.com}
\author[brin]{Edi Suprayoga\corref{cp}}
\ead{edi.suprayoga@brin.go.id}

\cortext[cp]{Corresponding author}
\address[brin]{Research Center for Quantum Physics, National Research and Innovation Agency (BRIN), South Tangerang 15314, Indonesia}

\begin{abstract}
Lead-free double perovskites are gaining attention for photovoltaic (PV) applications due to their long carrier lifetimes, tunable bandgaps, and low toxicity. Using first-principles calculations, we studied the structural, electronic and optical properties of \BX~($B=$ Bi, In; $X=$ Cl, Br, I). The cubic phase (space group Fm3m) was analyzed within the projector-augmented wave (PAW) method. Our calculations predict direct bandgaps of 1.9--1.2 eV for \BiX~and indirect bandgaps of 2.4--0.8 eV for \InX. Notably, the bandgap energy decreases with anion substitution from Cl to I, making these materials highly active in the near-infrared to visible light range. We reveal that \BiI~exhibits the highest optical absorption, with a peak value of $5\times10^5$ cm$^{-1}$ at an incident photon energy of 3 eV. Additionally, we evaluated the transport properties using the Boltzmann transport equations. The results indicate that \BiX~exhibit high electrical conductivity, reaching $8\times10^6$ S/m, and high electron mobility of 120 cm$^2/$V.s. PV performance analysis further reveals promising power conversion efficiencies (PCE) of up to 42\%, with \BiX~showing significantly higher PCE than \InX. These reports highlight the potential of \BiX~for advanced photovoltaic devices.
\end{abstract}

\begin{keyword}
  double perovskite\sep electronic properties\sep optical properties\sep photovoltaic.
\end{keyword}

\end{frontmatter}

\section{Introduction}
\label{sec:intro}

Perovskite photovoltaic (PV) have undergone significant development since the discovery of metal-halide perovskites, with their efficiency increasing from 3.8\% to 25.7\% in recent years \cite{liu2023recent}. Despite their advantageous properties, such as high carrier mobility, tunable bandgaps, low excitation energy, and strong light absorption, lead-based halide perovskites face critical challenges \cite{mcclure2016cs2agbix6,zeng2020highly,adnan2024electronic}. These include instability at room temperature, and the main issue is the toxic nature of lead, which poses a major barrier to commercialization. Therefore, researchers are exploring alternative all-inorganic perovskites that are nontoxic and lead-free \cite{liu2021lead,berri2023first}. Among these, lead-free double perovskites have emerged as a promising avenue for future development.


Previous studies have investigated Sn$^{2+}$ and Ge$^{2+}$ cations as potential substitutes for Pb$^{2+}$; However, these elements are easy to oxidate in ambient environments, leading to limitations in their practical utilization \cite{ghosh2019,stoumpos2013}. More importantly, compared to lead-based perovskite solar cells, the maximum PV power conversion efficiency (PCE) of tin-based devices has reached only 6.4\% \cite{noel2014}. An alternative approach to improve the potential of the perovskite family for solar applications is to replace Pb$^{2+}$ with relatively less toxic trivalent cations, such as Bi$^{3+}$ and Sb$^{3+}$. This substitution results in a new structural formula, $A_3M_2X_9$ (where $A=$ Cs; $M=$ Bi, Sb; and $X=$ Cl, Br, I), known as vacancy-ordered perovskites \cite{akkerman2020}. However, this structure is less suitable for optoelectronic applications due to its lower dimensional connectivity of metal halide octahedra compared to typical $ABX_3$ perovskite, which leads to reduced electronic mobility and poorer charge transport properties \cite{fan2020,volonakis2017}.


Another approach to the production of environmentally friendly halide double perovskites (HDPs) involves combining trivalent Bi$^{3+}$ with a monovalent cation, forming the elpasolite structure with the general formula $A_2MBX_6$ (where $A$ and $M=$ Rb, Cs, Tl, Ag; $B=$ In, Al, Bi, Ga; and $X=$ Cl, Br, I) \cite{giustino2016}. This structure has been synthesized in more than 350 compositions, demonstrating its versatility \cite{hoye2018}. Among these, Cs$_2$AgBiBr$_6$ has been extensively studied and demonstrates promising properties, including a long carrier lifetime of 1 $\mu$s, a suitable bandgap of 1.95--2.19 eV, moderate charge-carrier mobility, low effective carrier mass, and high stability with low toxicity \cite{lei2021lead}. In particular, the optimized PCE of Cs$_2$AgBiBr$_6$-based devices has improved from 1.5\% to 6.27\%, while also exhibiting bright emission characteristics \cite{albalawi2022}. Volonakis \textit{et al.} reported the first thermally and mechanically stable double perovskite, Cs$_2$AgInCl$_6$, with a direct bandgap of 2.5--3.3 eV and lattice parameters ranging from 10.469--10.481 \AA~\cite{wang2021}. Furthermore, studies of density functional theory (DFT) have revealed the significant role of Tl cations in shaping the edges of the valence and conduction band edges of Cs$_2$KTl$X_6$ ($X=$ Cl, Br, I), highlighting their influence on the optical, electronic, and transport properties of the material \cite{Giannozzi2009}.

The latest research, therefore, tends to encourage the development of potential candidates on the basis of Tl, Bi, and In halides for sustainable energy applications. In this study, we present a comprehensive investigation of the structural, electronic and PV performance of double perovskite Cs$_2$Tl$BX_6$ ($B=$ Bi, In; $X=$ Cl, Br, I). Using first-principles DFT calculations with the generalized gradient approximation (GGA) and Heyd-Scuseria-Ernzerhof (HSE) hybrid functional, we systematically analyzed the optoelectronic properties of these materials. To evaluate their potential for photovoltaic applications, we performed numerical simulations using \textsc{SCAPS-1D}, examining key performance metrics such as band diagram, open circuit voltage, short-circuit current density, fill factor (FF), and power conversion efficiency. Our theoretical study would provide a foundation for the future development of high-performance, lead-free solar cells based on halide double perovskites.

\section{Computational Details}
\label{sec:method}
The structural, electronic and optical properties of double perovskite \BX~were thoroughly investigated using density functional theory (DFT) as implemented in the \textsc{Quantum Espresso} package \cite{Giannozzi2009}. We used projector augmented wave (PAW) pseudopotentials to describe the core and valence electrons as follows: 2 for Cs ([Xe]$1s$), 1 for Tl ([Xe] $4f^{14} 5d^{10} 6s^2 6p^1$), 1 for Bi ([Xe] $4f^{14} 5d^{10} 6s^2 6p^3$) or In ([Kr] $4d^{10} 5s^2 5p^1$), and 6 for Cl ([Ne] $3s^2 3p^5$), Br ([Ar] $3d^{10} 4s^2 4p^5$), or I ([Kr] $4d^{10} 5s^2 5p^5$). The exchange-correlation potential was determined using the generalized gradient approximation (GGA) method developed by Perdew–Burke–Ernzerhof (PBE) \cite{PBE} since this approximation can accurately estimate the properties. The plane basis set was employed with the cutoff energy and charge density of 60 and 720 Ry, respectively. The $k$-point meshes were set to $6\times6\times6$ in the Monkhorst–Pack scheme. The denser $k$-point mesh of $30\times30\times30$  was generated to estimate the electronic and optical properties. Due to the presence of incredibly heavy atoms, spin-orbit coupling (SOC) and fully relativistic effects must unavoidably be considered. We also presented the bandgap correction using a DFT with a Heyd-Scuseria-Ernzerhof (HSE) \cite{Heyd2003} hybrid functional for comparison. The mixing parameter and the screen parameter were the two key parameters that contributed to setting the HSE functional. The exact exchange functional between the PBE and Hartree-Fock functionals can be modified using these parameters. The values of the mixing parameter and the screen parameter for a standard HSE calculation are 0.25 and 0.2, respectively, which correspond to the HSE06 functional.

From the data on the electronic structure ($E_{i,k}$), we can calculate the electrical conductivity ($\sigma$) through the Boltzmann transport equation, expressed as
\begin{equation}\label{eq:sigma}
\sigma = -\frac{q^2}{NV} \sum_{i,\textbf{k}}{\tau \nu_{i,\textbf{k}}^2 \frac{\partial f_{i,\textbf{k}}}{\partial E_{i,\textbf{k}}}},
\end{equation}
where $q=\pm 1.602 \times 10^{-19}$ C is the carrier charge for the hole and the electron, respectively, $N$ is the number of $k$-points within the Brillouin zone, $V$ is the volume of a unit cell and $\tau$ is the electron relaxation time. Here, $f_{i,k}$ is the Fermi-Dirac distribution function and $\nu_{i,\textbf{k}}= \partial E_{i,\textbf{k}} / \partial \textbf{k}$ is the electron group velocity as a function of band index $i$ and wave vector $\textbf{k}$. 

The optical properties were investigated by computing the complex dielectric function $\varepsilon(\omega)=\varepsilon_1 + i \varepsilon_2$. The real parts of the dielectric function ($\varepsilon_1$) describe the electric polarization responses of the material under an external field. Hence, the imaginary parts of the dielectric function ($\varepsilon_2$) represent the absorptive behavior and the loss tangent. The Kramers–Kronig equation was used to determine both parts, which is expressed below \cite{fischer1992general}.
\begin{equation}\label{eq:reald}
\mathrm{\varepsilon_1}(\omega) = 1+\frac{2}{(\pi)}P \int_{0}^{\infty}{\frac{\omega'
\varepsilon_2 (\omega')}{\omega'^2-\omega^2}d\omega' }
\end{equation} 
\begin{equation}\label{eq:imaganaryd}
\mathrm{\varepsilon_2}(\omega) = \frac{4\pi e^2}{m^2 \omega^2} \sum_{ij} \int{\langle i |M|j\rangle I^2 f_i (1- f_i) \delta (E_f - E_i - \omega) d^3 k }
\end{equation}
Here, $\omega$, $P$, $e$, and $m$ are the frequency of incident photons, the integral primary value, electronic charge, and mass of the free electron, respectively. Meanwhile, $M$, $f_i$, and $E$ denote the dipole matrix, the Fermi distribution, and the free-electron energy, respectively.

To understand the fundamentals of HDP solar cells, numerical modeling, such as Solar Cell Capacitance Simulator One Dimension (\textsc{SCAPS-1D}) software \cite{SCAPS}, was used. The one-dimensional solar cells can be solved with three coupled differential equations, for instance, the Poisson equation, the continuity equation for electrons, and holes. The Poisson equation describes the correlation between the charge density distribution and electrostatic potential, as given by:
\begin{equation}\label{eq:poisson}
   \begin{split}
    \frac{d^2}{dx^2} \psi(x) &= \frac{q}{\varepsilon_0 \varepsilon_r} \left[p(x)-n(x)+N_D-N_A+ \rho_p - \rho_n\right]
   \end{split}
\end{equation}
where $n(p)$, $N_D$($N_A$), $\rho_n$ ($\rho_p$), and $q$ represent the density of the electron (hole), the ionized donor (acceptor), the trapped electron (hole), and the elementary electronic charge, respectively.

After the potential is estimated using the Poisson equation, the concentration of the electron-hole is determined by the continuity equation. The steady-state electron and hole continuity is written by \eqref{eq:continu1} and \eqref{eq:continu2}, respectively.
\begin{equation}\label{eq:continu1}
   \begin{split}
    \frac{dn}{dt} &= \frac{1}{q} \frac{\partial J_n}{\partial x}+\left(G_n-R_n\right)
   \end{split}
\end{equation}
\begin{equation}\label{eq:continu2}
   \begin{split}
    \frac{dp}{dt} &= \frac{1}{q} \frac{\partial J_p}{\partial x}+\left(G_p-R_p\right)
   \end{split}
\end{equation}
Here, $J_n$ ($J_p$) represents for the current density, $G_n$ ($G_p$) for the generation rate, $R_n$ ($R_p$) for the recombination rate of electron (hole). The both current density is expressed by:
\begin{equation}\label{eq:jn}
 J_n = q \mu_n n \in + q D_n \partial_n
\end{equation}
\begin{equation}\label{eq:jn2}
 J_p = q \mu_p p \in + q D_p \partial_p
\end{equation}
where $\mu_n$ ($\mu_p$) is the mobility electron (hole) and $D_n$ ($D_p$) is the diffusion coefficient of the electron (hole), respectively.

Finally, the open circuit voltage ($V_\textrm{OC}$), the fill factor (FF), and the PCE can be calculated using equation \eqref{eq:voc}, \eqref{eq:ff}, and \eqref{eq:pce}, respectively.
\begin{equation}\label{eq:voc}
 V_\textrm{OC} = \frac{E_{\textrm{Fp}}-E_{\textrm{Fn}}}{q}
\end{equation}
\begin{equation}\label{eq:ff}
 \textrm{FF} = \frac{I_{\textrm{max}}\times V_{\textrm{max}}}{I_{\textrm{SC}}\times V_{\textrm{OC}}}
\end{equation}
\begin{equation}\label{eq:pce}
 \textrm{PCE} = \frac{\textrm{FF}\times V_{\textrm{OC}} \times I_{\textrm{SC}}}{P_{\textrm{in}}}
\end{equation}
Here, $E_{\textrm{Fp}}$ and $E_{\textrm{Fn}}$ represent the quasi-Fermi level of free electron and hole, respectively. Furthermore, $I_{\textrm{max}}$, $V_{\textrm{max}}$, $I_{\textrm{SC}}$, and $P_{\textrm{in}}$ denote the maximum current, maximum voltage, short circuit current, and input power, respectively.

\section{Results and Discussion}
\label{sec:result}

\subsection{Structural stability}
\label{sub:struc}

\begin{figure}[tb]
    \centering
    \includegraphics[width=8cm]{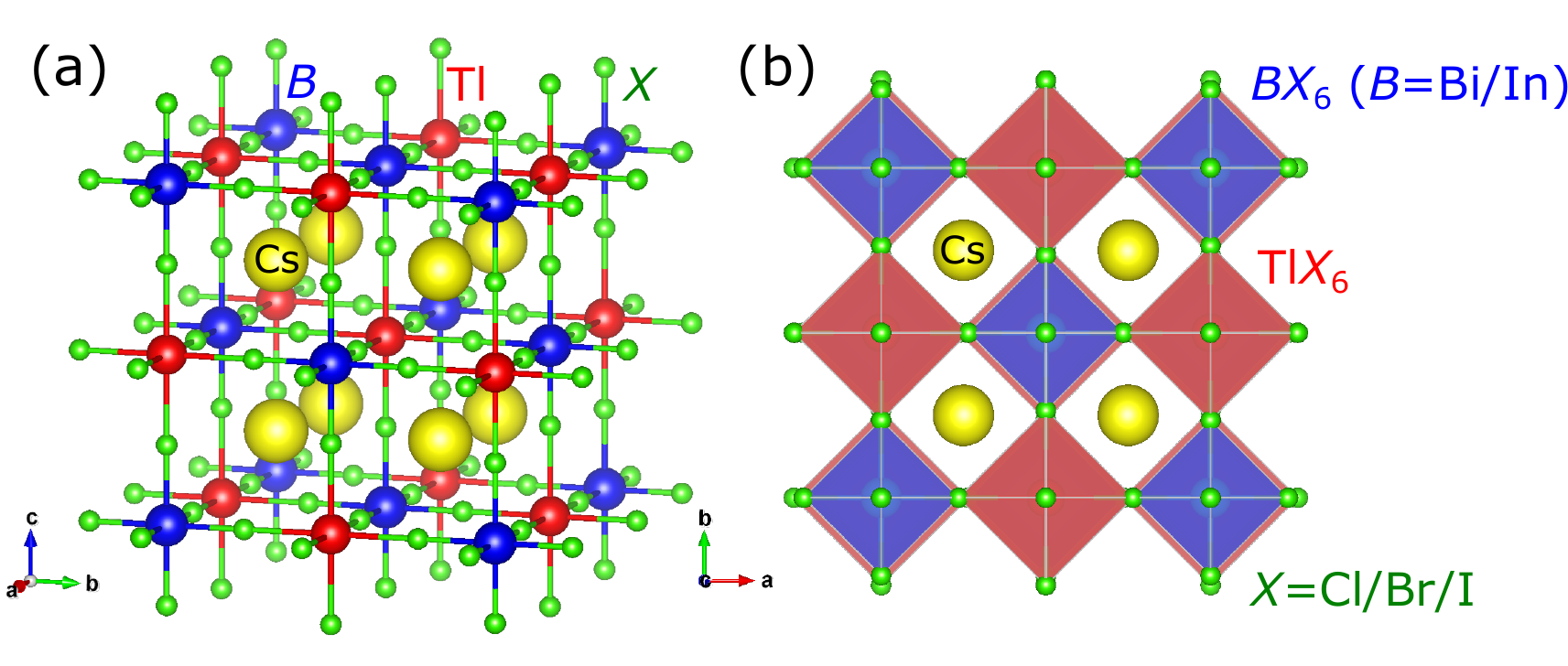}
    \caption{Conventional structure of double perovskite Cs$_2$Tl$BX_6$ with cubic symmetry seen from (a) the best angle and (b) top view.}
    \label{Fig:stucture}
\end{figure}

\begin{figure}[tb]
    \centering
    \includegraphics[width=8cm]{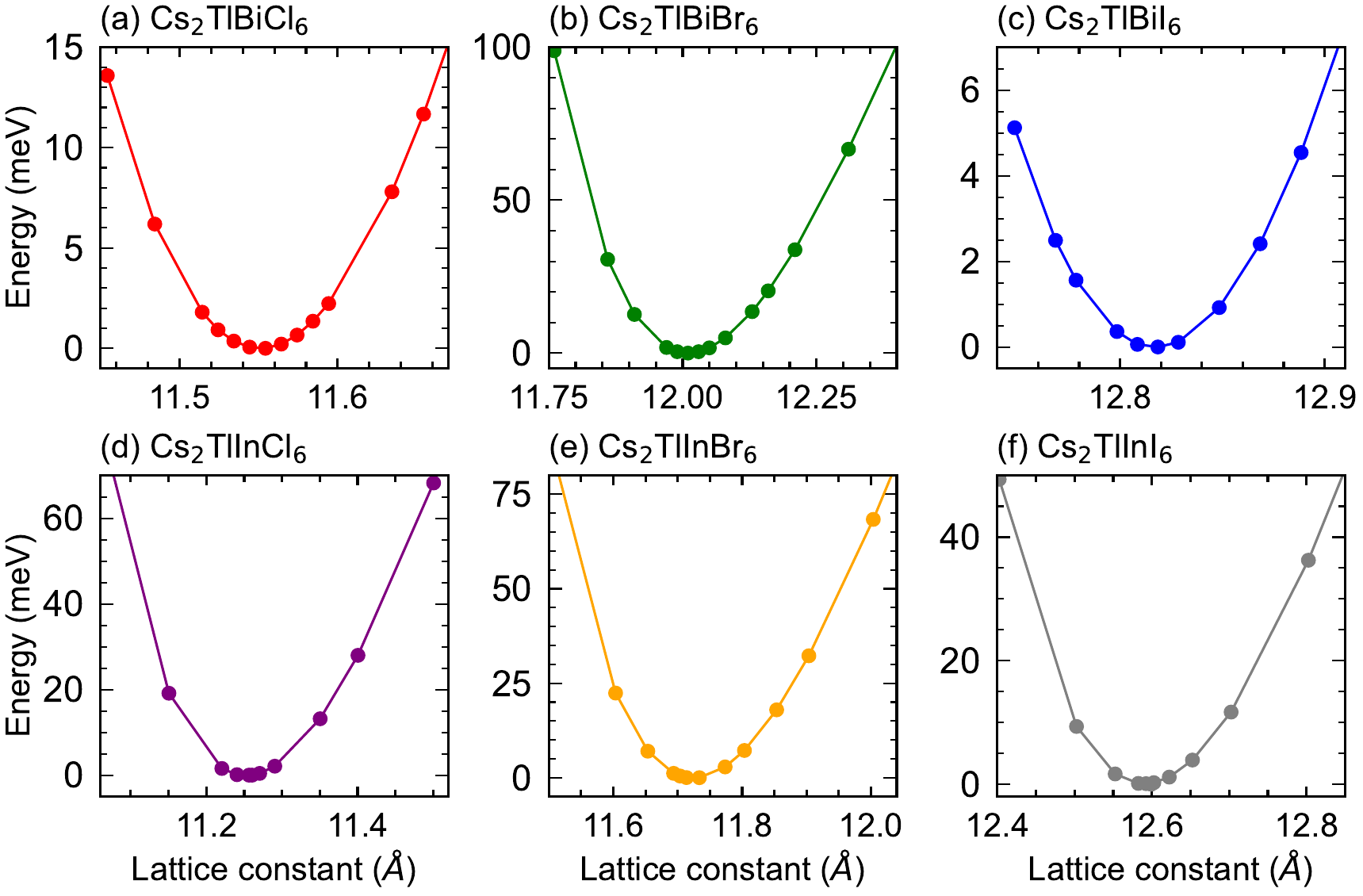}
    \caption{Energy variation as a function of the lattice parameter for double perovskite Cs$_2$Tl$BX_6$.}
    \label{Fig:stucturestability}
\end{figure}

We generated structures of \BX~compounds with space group Fm-3m (No.225) which contain both octahedra Tl$X_6$ and $BX_6$ as illustrated in Fig. \ref{Fig:stucture}. In these crystal structures, Cs, (Tl, Bi/In), and $X$ occupy the Wyckoff positions $8c$, $4(a, b)$, and $24e$, respectively, with the corresponding oxidation states of $+1$, $+1$, $+3$, and $-1$. Unlike common perovskite, HDP unit cells feature octahedra positioned not only at the corners but also at the body-centered sites, resulting in an expanded unit cell \cite{roknuzzaman2017towards}. We optimized the cubic structures of all compounds using the GGA-PBE functional, and the total energy ($E_\mathrm{total}$) as a function of the lattice constant ($a$) is shown in Fig. \ref{Fig:stucturestability}. It is confirmed that all materials have a stable configuration that was found to have minimum energy. The minimum $E_\mathrm{total}$ corresponds to lattice constants of approximately 11.55 \AA, 12.01 \AA, 12.82 \AA, 11.26 \AA, 11.73 \AA, and 12.6 \AA~for \BX~($B=$ Bi; In and $X=$ Cl; Br; I), respectively, as presented in Table~\ref{tab:param1}. These results demonstrate that the lattice constant increases with the atomic size of the halogen, reflecting the influence of halogen size on the structural properties of the materials.

The structural stability of the HDP \BX~is crucial for developing reliable materials for solar cells and other optoelectronic applications. To assess the stability of the structure, we employed the Goldschmidt tolerance factor ($t_G$) and the octahedral ratio ($t_{O}$) formula. The Goldschmidt tolerance factor evaluates the size effects of HDP crystal structures, whereas the octahedral factor takes into account overall stability, which can be written as \cite{cai2019high}:
\begin{equation}\label{eq:t}
   \begin{split}
    t_G &= \frac{r_{A}+r_X}{\sqrt{2} \left(\frac{r_B+r_{B'}}{2}\right)+r_X}, \\
    t_O &= (r_B+r_{B'})/2r_{X}
   \end{split}
\end{equation}
where $r_A$, $r_B$, $r_{B'}$, and $r_X$ are average ionic radii of  Cs, Tl, Bi/In, and $X$ atoms, respectively. The ionic radii used in our calculations are derived from established scientific references \cite{shannon1976revised,li2008formability}. The ideal Goldschmidt tolerance factor is in the range between $0.8$ and $1.00$, while the stable octahedral ratio must be equal to or greater than $0.41$. Table \ref{tab:param1} summarizes the tolerance factor and octahedral ratio for all compounds. Although the tolerance factor suggests a nearly stable cubic structure for these materials, the octahedral ratios for all the compounds lie within the stable range, thereby confirming their structural stability.

To evaluate the dynamical stability of \BX~compounds, we performed phonon dispersion calculations using density functional perturbation theory, as presented in Fig. S7 (Supplementary Materials). The presence of imaginary frequencies (negative modes) in the phonon branches indicates dynamical instabilities in certain Cs$_2$Tl$BX_6$ compounds, particularly those containing Bi.
These theoretical predictions align with experimental challenges in synthesizing stable Cs$_2$TlBi$X_6$ compounds, where Tl$^+$ oxidation to Tl$^{3+}$ under ambient conditions leads to decomposition into Cs$_3$Tl$_2$Cl$_9$ + BiCl$_3$ \cite{Fuxiang2023}.
Among the materials studied, only Cs$_2$TlInCl$_6$ exhibits fully real frequencies throughout the entire Brillouin zone, confirming its thermodynamical stability within the harmonic approximation.
However, further experimental validation is needed to confirm the stability and viability of these materials for photovoltaic applications.

\begin{table*}
  \centering
    \begin{tabular}{ccccccc}
       \hline
       Parameters & \BiCl & \BiBr & \BiI & \InCl & \InBr & \InI \\
       \hline
       Lattice constant (\AA) & 11.554 & 12.009 & 12.818 & 11.256 & 11.733 & 12.599\\
       Volume (\AA$^3$) & 1542.401 & 1731.891 & 2106.012 & 1426.107 & 1615.203 & 1999.899\\
       Cell density & 3.974 & 4.531 & 4.702 & 3.809 & 4.414 & 4.582\\
       Tolerance factor & 0.765 & 0.763 & 0.759 & 0.794 & 0.785 & 0.782\\
       Octahedral factor & 0.776 & 0.716 & 0.638 & 0.713 & 0.658 & 0.586\\
       \hline
    \end{tabular}
    \caption{The calculated structural parameters including lattice constant, volume, cell density, Goldschmidt tolerance factor and octahedral factor for double perovskites \BX.}
    \label{tab:param1}
\end{table*}

\subsection{Electronic properties}
\label{sub:elect}
\begin{figure*}[ht!]
    \centering
    \includegraphics[width=\textwidth]{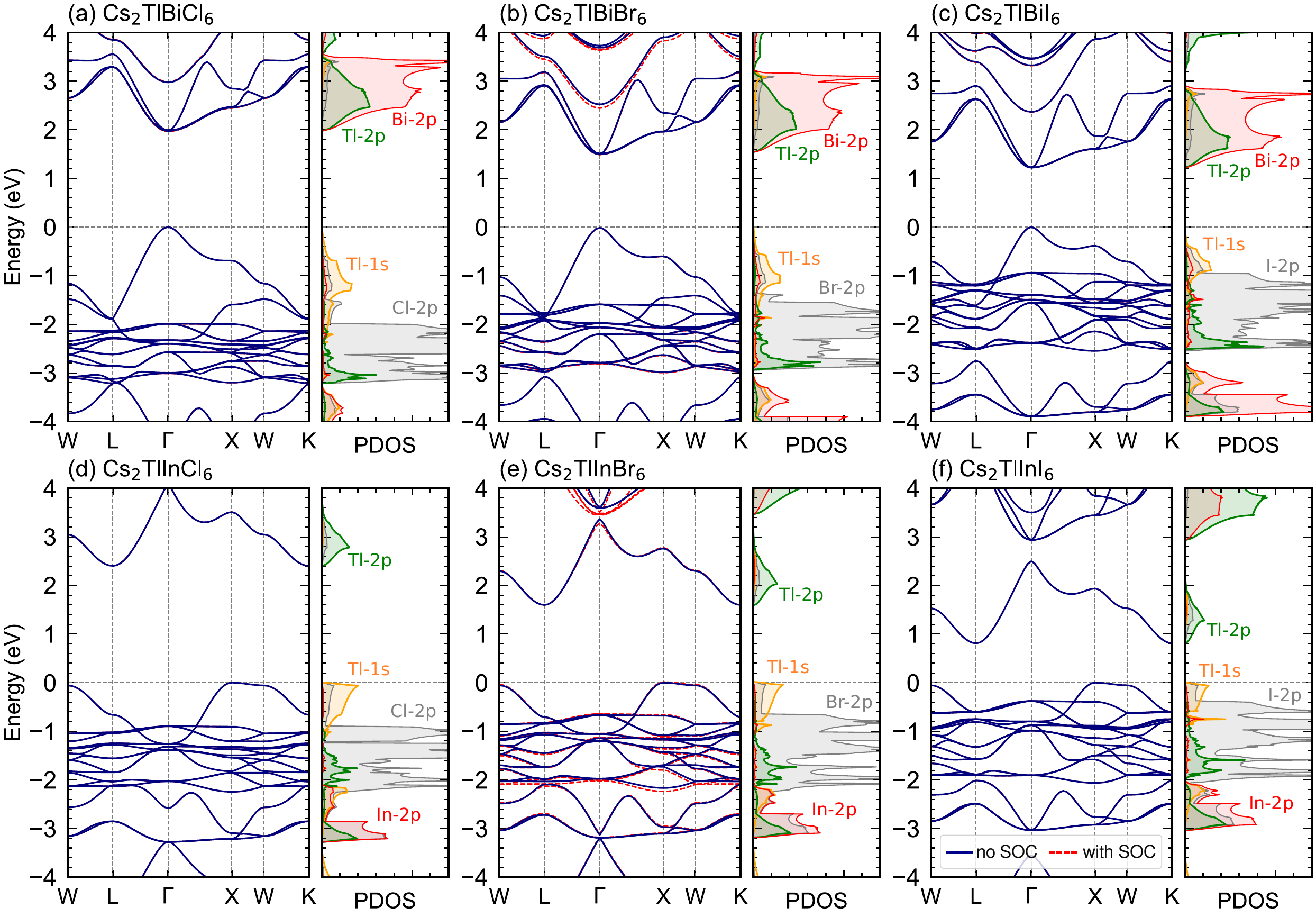}
    \caption{Electronic band structure and projected density of state of double perovskites Cs$_2$Tl$BX_6$. The blue solid line and red dash line in the band structure plots represent the calculation without and with SOC, respectively.}
    \label{Fig:bands}
\end{figure*}

 Electron dispersion is a key factor in determining the classification of materials, such as insulators, semiconductors, semi-metals or metals \cite{ghosh2024strain}. Moreover, the physical properties of perovskite materials are closely related to their electronic band structure, where charge transfer and optical properties are influenced by the bandgap \cite{rahman2023investigation}. To investigate the correlation between composition, structure, and electronic properties, we calculated the band structures of all compounds using the \textsc{Quantum Espresso} package. In this work, we used the GGA-PBE exchange-correlation functional, both with and without SOC, to determine the electronic band structure, as shown in Fig. \ref{Fig:bands}. The SOC calculation may obtain accurate bandgap values close to the experimental values \cite{deng2016exploring,qi2024exploring}. Our calculations reveal that the HDP band structures exhibit bandgaps of 1.975 eV, 1.518 eV, and 1.225 eV for \BiCl, \BiBr, and \BiI, respectively. Furthermore, the bandgaps for \InCl, \InBr, and \InI~are 2.402 eV, 1.599 eV, and 0.812 eV, respectively, all at the $\Gamma$-point in the Brillouin zone. These results highlight the tunability of the electronic properties based on the composition of the materials.

The bandgap is a critical factor in determining the light absorption efficiency of optoelectronic devices \cite{nair2019cs2tlbii6}. Previous studies have demonstrated that bandgap values can be effectively tuned through halogen substitution \cite{li2008formability}. Prior studies on Cs$_2$AlAg$X_6$ show that it has a direct bandgap, with a value of 2.41--3.05 eV determined using HSE06 \cite{iqbal2025intrinsic}. 
In our calculations, we observe that the bandgap energy decreases slightly as the anion changes from Cl to I, resulting in more pronounced semiconducting behavior. Our results suggest that the HDP \BX~are well-suited for optoelectronic applications, as their bandgap values fall within the near-infrared and visible light spectrum energy ranges. Specifically, bismuth-based HDPs, \BiX~exhibit a direct bandgap at the $\Gamma$-point, which facilitates efficient electron transfer from the valence band (VB) to the conduction band (CB). However, indium-based HDPs, \InX~possess an indirect bandgap, with transitions occurring from the $X$-point on the VB to the $L$-point on the CB, making them less suitable for solar cell applications.

Table \ref{tab:bandgap} the calculated energy bandgaps for \BX~using the PBE and HSE06 functionals, both with and without SOC. Notably, there are currently no experimental data available in the literature for the remaining compounds. Our calculations reveal a consistent trend across all DFT techniques, with the bandgap decreasing in the order $E_g$(Cl) $>$ $E_g$(Br) $>$ $E_g$(I). Furthermore, the bandgaps calculated using the PBE and HSE06 functionals exhibit different tendencies. In fact, the HSE06 functional, which accounts for the self-energy of a many-body electron system and considers significant atomic interactions, provides more accurate results \cite{kaewmeechai2021dft}. When SOC is included in the calculations, the energy bandgaps show a slight shift of approximately 0.02 eV at the $\Gamma$-point, as illustrated by the dashed line in Fig. \ref{Fig:bands}. 

The density of states (DOS) and the partial density of states (PDOS) have been analyzed to understand the contributions of individual orbital states to the energy bands. As illustrated in Fig. \ref{Fig:bands}, Cs, Bi, Tl, In, and the halogens significantly influence the valence and conduction bands. The orbitals Tl-$1s$, Cl-$2p$, Br-$2p$, and I-$2p$ contribute predominantly to the DOS near the valence band maximum (VBM), while the orbitals Tl-$2p$, Bi-$2p$, and In-$2p$ dominate the DOS near the conduction band minimum (CBM). These results indicate that the electronic properties of HDPs are strongly influenced by the interactions between Tl with $B$ and $X$ atoms around the Fermi level. Notably, the $p$-states of \BX~($X =$ Cl, Br, I) provide the highest contribution to the carrier distribution below the Fermi level in the valence band. 

We also calculated the effective electron mass ($m_e$) and the hole ($m_h$) relative to the free electron mass ($m_0$) by analyzing the curvature of the VBM and CBM, respectively. The effective masses were determined using the following equation \cite{feng2017double}:
\begin{equation}\label{eq:effective}
   \begin{split}
    \frac{1}{m} &= \frac{1}{\hbar} \frac{\delta^2E}{\delta k^2}
   \end{split}
\end{equation}
The computed values of $m_e/m_0$ and $m_h/m_0$ are presented in Table \ref{tab:bandgap}. The VBM curves of \InX~are richer and flatter compared to those of \BiX, indicating that the \InX~compounds have a greater effective electron mass. Among the studied materials, \InCl~shows the largest effective electron and hole mass, with values of 1.1 and 0.5, respectively. In contrast, the \BiI~exhibits smallest effective electron and hole mass, with values of 0.1 and 0.3, respectively. A smaller effective mass generally leads to higher carrier mobility \cite{mcclure2016cs2agbix6}, suggesting that \BiX~compounds are predicted having excellent carrier mobility for photoinduced carriers.

\begin{table*}
  \centering
    \begin{tabular}{cccccccc}
       \hline
       \multirow{2}{*}{Materials} & $E_g$/eV & $E_g$/eV & $E_g$/eV & $E_g$/eV & 
       \multirow{2}{*}{$m_h/m_0$} & \multirow{2}{*}{$m_e/m_0$} \\
                        & (PBE) & (PBE+SOC) & (HSE06) & (HSE06+SOC) &  &  &  \\
       \hline
         \BiCl & 1.975 & 1.964 & 2.047 & 2.047 & 0.273 & 0.403 \\
         \BiBr & 1.518 & 1.506 & 1.527 & 1.523 & 0.221 & 0.361 \\
         \BiI  & 1.225 & 1.221 & 1.213 & 1.213 & 0.187 & 0.346 \\
         \InCl & 2.402 & 2.403 & 2.421 & 2.421 & 0.564 & 1.175 \\
         \InBr & 1.599 & 1.586 & 1.589 & 1.589 & 0.468 & 1.027 \\
         \InI  & 0.812 & 0.809 & 0.804 & 0.803 & 0.403 & 0.995 \\
       \hline
    \end{tabular}
    \caption{Comparison of the computed energy gap ($E_g$) using PBE, SOC, and HSE06 methods, along with the calculated electron (hole) effective mass, $m_e (m_0)$, determined from PBE+SOC.}
    \label{tab:bandgap}
\end{table*}

\subsection{Optical properties}
\label{sub:optic}

\begin{figure*}[ht!]
    \centering
    \includegraphics[width=\textwidth]{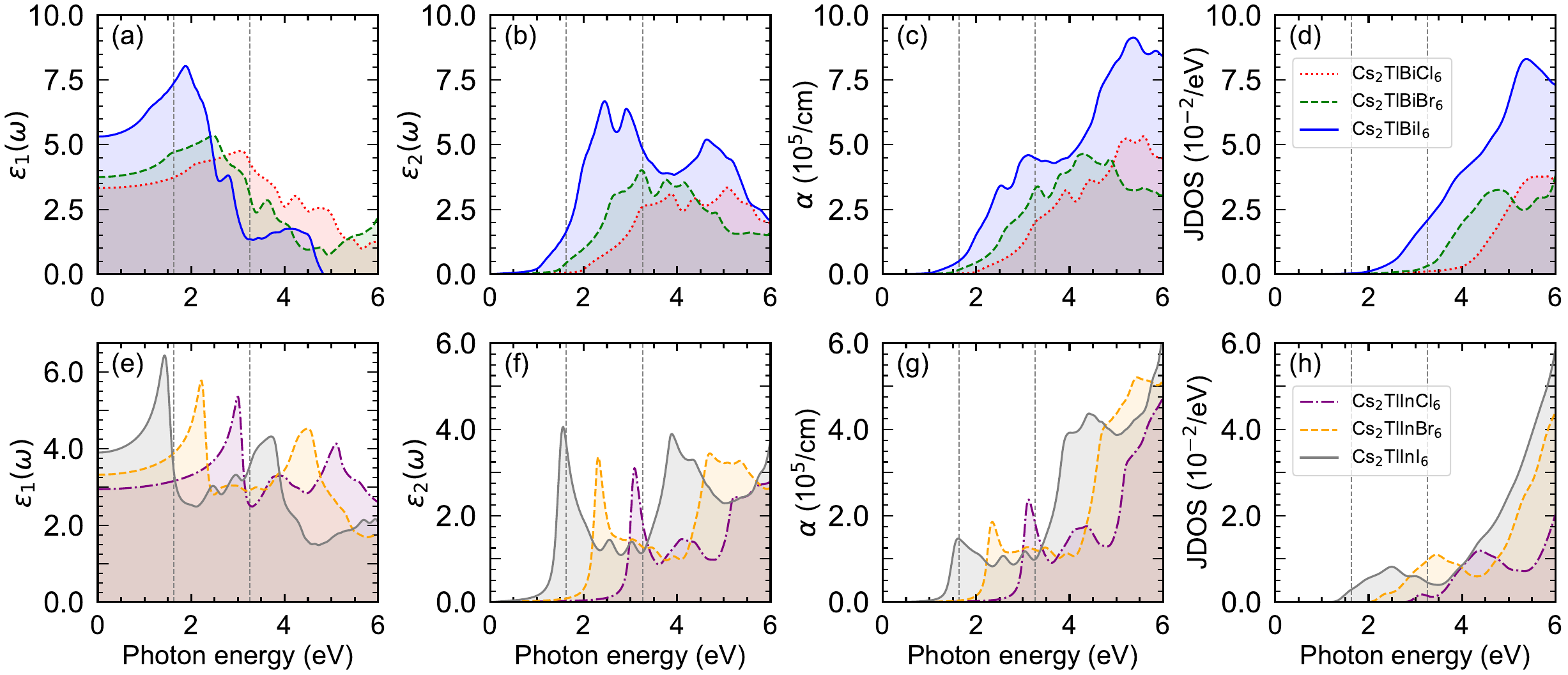}
    \caption{Optical properties of double perovskites (a--d) Cs$_2$TlBi$X_6$ and (e--h) Cs$_2$TlIn$X_6$. (a, e) Real part and (b, f) imaginary part of the dielectric tensor. (c, g) Absorption coefficient and (d, h) joint density of states as a function of photon energy. The vertical dashed lines represent the visible light frequency range.}
    \label{Fig:optic}
\end{figure*} 

To investigate the optical properties of \BX, we calculated the real ($\varepsilon_1$) and imaginary ($\varepsilon_2$) parts of the dielectric function, the absorption coefficient ($\alpha$), and the joint density of states (JDOS) using the time-dependent density functional theory (TDDFT) method. The frequency-dependent dielectric function was first examined, with the real parts plotted as a function of photon energy in Figs. \ref{Fig:optic}(a) and \ref{Fig:optic}(e) for \BiX~and \InX, respectively. At zero (no photon) energy, the static values of $\varepsilon_1$ are 5.4 for \BiI~and 3.8 for \InI, which are higher than the corresponding values for other analogs. This is indicated by the strong polarization and leads to increased conductivity which corresponds to the atomic mass of the anion. The Pens model further explains the correlation between the static values and the energy gap. The $\varepsilon_1$ exhibit maximum values at resonance frequencies corresponding to photon energies of 1.95 eV, 2.25 eV, and 2.5 eV for \BiX~($X=$ I, Br, and Cl), respectively. Once resonance was increased, the peaks began to rapidly drop to their minimal values. Furthermore, the spectrum of \InI~present slightly differs with two maximum values of resonance frequencies at 1.4 eV/3.6 eV, 2.2 eV/4.5 eV, and 3.1 eV/5.2 eV for \InX~($X=$ I, Br, and Cl), respectively, consistent with the typical behavior of materials with an indirect bandgap. 

Figures \ref{Fig:optic}(b) and \ref{Fig:optic}(f) show the imaginary part of the dielectric function as a function of the photon energy. The peak values of $\varepsilon_2$ lie within the visible light spectrum (1.5 to 3.0 eV). This range aligns well with the energy requirements for efficient light absorption in optoelectronic applications such as solar cells and photodetectors.

The optical absorption, which quantifies the amount of light absorbed by a material at a given wavelength, provides further insight into its potential for efficient solar energy conversion \cite{mouna2024structural, wang2021absolute}. The optical absorption coefficient $\alpha(\omega)$ is derived from the real and imaginary parts of the dielectric functions as follows. 
\begin{equation}\label{eq:alpha}
\alpha(\omega) = \frac{2\omega}{c} \sqrt{\frac{|\varepsilon(\omega)| - \mathrm{Re}\left(\varepsilon(\omega)\right)}{2}},
\end{equation}
where $c$ represents the speed of light in a vacuum. The key features of the optical devices are highlighted in the optical absorption behavior, as seen in Figs. \ref{Fig:optic}(c) and \ref{Fig:optic}(g). We confirm that the \BX~exhibits isotropic behavior, with $\alpha_{xx} = \alpha_{yy} = \alpha_{zz}$, due to its cubic crystal structure. The computed absorption coefficients reveal that \BiX~compounds exhibits higher optical absorption compared to \InX, with \BiI~showing the highest peak value of $5\times10^5$ cm$^{-1}$ at 3 eV. This enhanced absorption is attributed to the lower and direct bandgap energy. In addition, the absorption spectra for all compounds begin in the range of 1 to 2 eV, demonstrating strong absorption across the near-infrared (IR) to visible-light wavelength. 

The absorption coefficient becomes large when there is a high number of states for a pair of initial and final states at a specific energy difference $E$. The total number of electron-hole pairs generated at a specific excitation energy can be described using the joint density of states (JDOS) \cite{hedley2018hot}, which is defined as
\begin{equation}\label{eq:jdos}
\mathrm{JDOS}(E) = \frac{V}{(2\pi)^3} \int{\delta[E_c(\textbf{k})-E_v(\textbf{k})-E]\ d\textbf{k} },
\end{equation}
where $E_c(\textbf{k})$ and $E_v(\textbf{k})$ represent the energy dispersion of the conduction and valence bands, respectively. We assume that the photon wave vector is much smaller than the reciprocal lattice vector, allowing us to neglect it in terms of momentum conservation. This assumption is valid for photon energies $\hbar\omega < 3$ eV. In Figures \ref{Fig:optic}(d) and \ref{Fig:optic}(h), we present JDOS for \BiX~and \InX, respectively. The results indicate that the broad lowest energy peak of JDOS does not correspond to the energy gap and absorption coefficient plots. Furthermore, it is widely accepted that each energy transition is broadened using a Lorentzian or Gaussian function to characterize excitonic absorption. The JDOS intensity of \InX~is significantly higher than \BiX, which is attributed to in direct bandgap of \InX~and mediated by amount of electronic excited state. However, the JDOS peak shifts to lower energies as the halogen changes from Cl to I, due to the contribution of $2p$ orbitals of the halide in the valence band. 

\begin{figure*}[t]
    \centering
    \includegraphics[width=\textwidth]{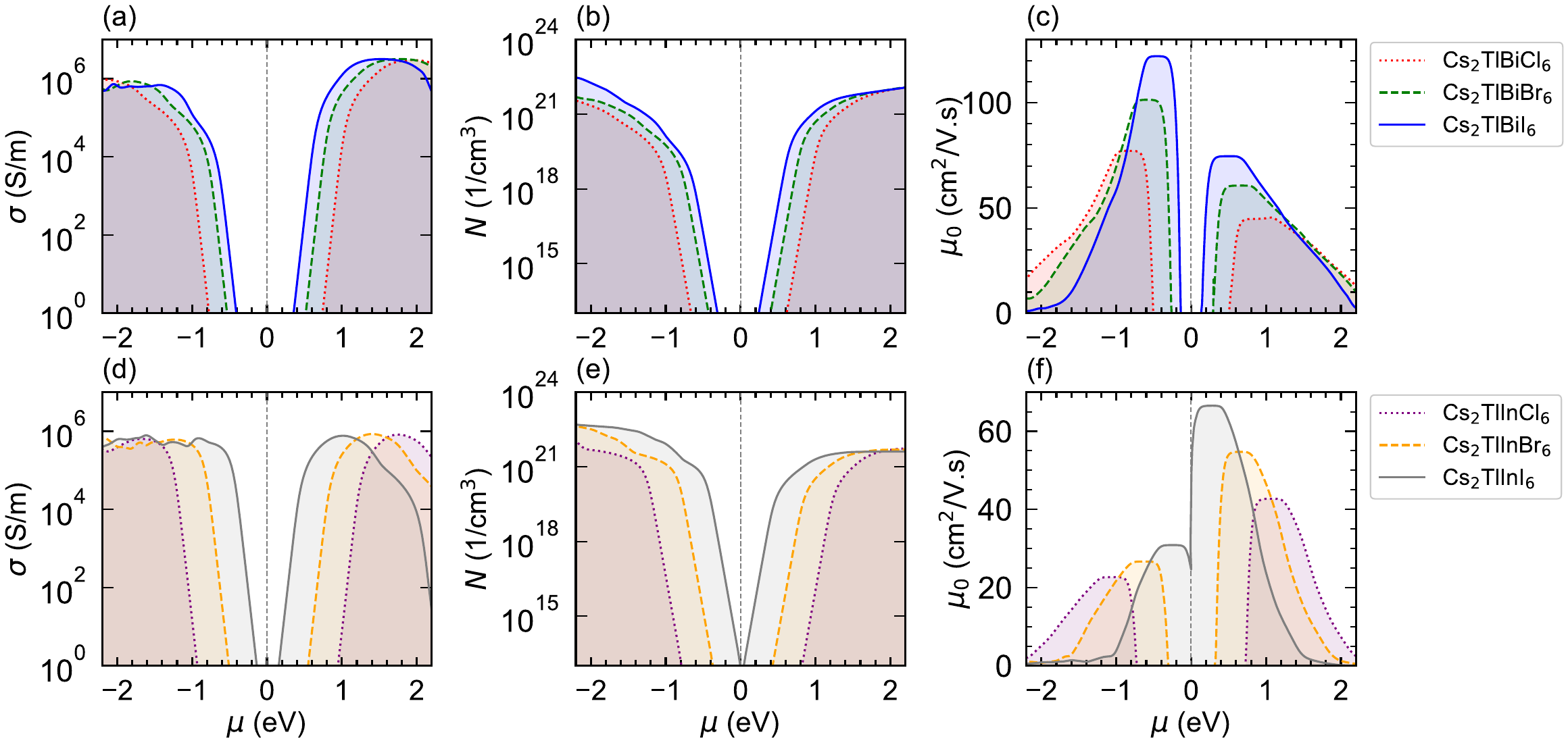}
    \caption{Electronic transport properties of (a--c) Cs$_2$TlBi$X_6$ and (d--f) Cs$_2$TlIn$X_6$ double perovskites. (a, d) Electrical conductivity, (b, e) carrier concentration, and (c, f) electron (hole) mobility as a function of chemical potential. The charge neutrality in this figure is positioned at the middle of the bandgap.}
    \label{Fig:boltz}
\end{figure*}

\subsection{Transport properties}
\label{sub:transport}
Charge transport properties are crucial parameters for determining the efficiency of waste heat into electrical energy \cite{li2016thermoelectric}. We calculated transport properties, including electrical conductivity ($\sigma$), carrier concentration ($N$), and electron (hole) mobility ($\mu_0$) as a function of chemical potential ($\mu$), as shown in Fig. \ref{Fig:boltz}. These properties were computed using the \textsc{BoltzTraP2} code \cite{Madsen2018}, which is based on the classical Boltzmann transport equation (BTE). A constant relaxation time of $\tau=1$ fs was employed, consistent with the moderately dirty regime theory \cite{DirtyRegime}, where the conductivity is expected to range from $10^5$ to $10^7$ S/m. In all cases, the charge carrier that contributes to electrical conductivity and mobility below $\mu=0$ eV are holes, while above $\mu=0$ eV are electrons. The calculations were performed at a constant temperature of 300 K, reflecting typical operating conditions for optoelectronic devices. 

The electrical conductivity behavior was analyzed to determine the trend of electron flow. Figures \ref{Fig:boltz}(a) and \ref{Fig:boltz}(d) show that the electrical conductivity of \BiX~and \InX~reach its peak values of approximately $8\times10^5$ S/m and $2\times10^6$ S/m, respectively. Notably, the conductivity values for holes are significantly lower than those for electrons. The effective masses of electrons and holes play a critical role in determining electrical conductivity, as they influence the mobility of charge carriers in perovskite solar cells. Consequently, the lower effective mass of the charge carriers in \BiX~results in higher conductivity compared to \InX. In addition, all compounds exhibit a similar trend in conductivity, with slight shift in the peak position as a result of halogen substitution.

Figures \ref{Fig:boltz}(b) and \ref{Fig:boltz}(e) illustrate the carrier concentrations of all compounds, which exhibit a similar trend with the highest value of approximately $10^{22}$ cm$^{-3}$ for \BiI. The higher carrier concentration in \BiI~has a favorable impact on carrier mobility. Additionally, the trend in carrier concentration shifts toward zero chemical potential as the halogen radius increases, which correlates with the observed decrease in bandgap energy. Furthermore, we can directly derive the carrier mobility ($\mu_0$) from the electrical conductivity as
\begin{equation}\label{eq:mu_e}
    \mu_0= \frac{\sigma}{n_e} q 
\end{equation}
where $n_e$ is the carrier concentration. 
We consider a constant scattering rate that assumes an isotropic and elastic scattering mechanism, independent of temperature and electron energy. Although electron (hole) mobility values require more scattering models, this approach remains valid to compare the relative performance of different Cs$_2$Tl$BX_6$ compounds.
According to Figs. \ref{Fig:boltz}(c) and \ref{Fig:boltz}(f), the hole mobility of \ch{Cs_2TlBi\textit{X}_6} reaches a higher value of approximately $\mu_0=120$ cm$^2/$V.s, compared to 30 cm$^2/$V.s for \InX. The hole mobility of \BiX~has a bit higher value compared to \InX. This indicates \BiX~has high electron-hole mobility in the conduction and valence band. 
The high electron mobility enables faster extraction of photogenerated carriers before recombination occurs. According to Eqs. (\ref{eq:jn}) and (\ref{eq:jn2}), the electron mobility is proportional to the current density and further improved the fill factor and power conversion efficiency,  which are described in Eqs. (\ref{eq:ff}) and (\ref{eq:pce}).
Furthermore, the effect of halides showed an initial increase pattern for both \BiX~and \InX. We can conclude that the \BiX~are excellent materials for applications related to optoelectronics compared to other compounds.

\begin{table*}
  \centering
    \begin{tabular}{cccccc}
       \hline
       Material Properties & ITO~\cite{kumar2021theoretical,singh2024enhancing} & ZnO~\cite{hossain2023designLa} & CsTl$BX_6$ [this work]\\
       \hline
       Thickness (nm) & 500 & 30 & 300 \\
       Bandgap (eV) & 3.5 & 3.3 & 1.9 \\
       Electron affinity (eV) & 4 & 4 & 3.6 \\
       Dielectric permittivity (relative) & 9 & 9 & 8.1 \\
       CB effective DOS (cm$^{-3}$) & $2.2 \times 10^{18}$ & $3.7 \times 10^{19}$ & $2.6 \times 10^{18}$ \\
       VB effective DOS (cm$^{-3}$) & $1.8 \times 10^{19}$ & $1.8 \times 10^{19}$ & $7.5 \times 10^{18}$ \\
       Electron thermal velocity (cm s$^{-1}$) & $1.0 \times 10^7$ & $1.0 \times 10^7$ & $1.0 \times 10^7$ \\
       Hole thermal velocity (cm s$^{-1}$) & $1.0 \times 10^7$ & $1.0 \times 10^7$ & $1.0 \times 10^7$ \\
       Electron mobility (cm${^2}$ V$^{-1}$ s$^{-1}$) & 20 & 100 & 2 \\
       Hole mobility (cm$^2$ V$^{-1}$ s$^{-1}$) & 10 & 25 & 2 \\
       Donor density (cm$^{-3}$) & $1.0 \times 10^{19}$ & $1.0 \times 10^{18}$ & - \\
       Acceptor density (cm$^{-3}$) & - & - & $1.0 \times 10^{15}$ \\
       $N_t$ (cm$^{-3}$) & $1.0 \times 10^{15}$ & $1.0 \times 10^{15}$ & $1.0 \times 10^{15}$ \\
       \hline
    \end{tabular}
    \caption{The physical parameters set in simulating double perovskites Cs$_2$Tl$BX_6$.}
    \label{tab:param2}
\end{table*}

\subsection{Photovoltaic performances}
\label{sub:pv}
\begin{figure*}[ht]
    \centering
    \includegraphics[width=\textwidth]{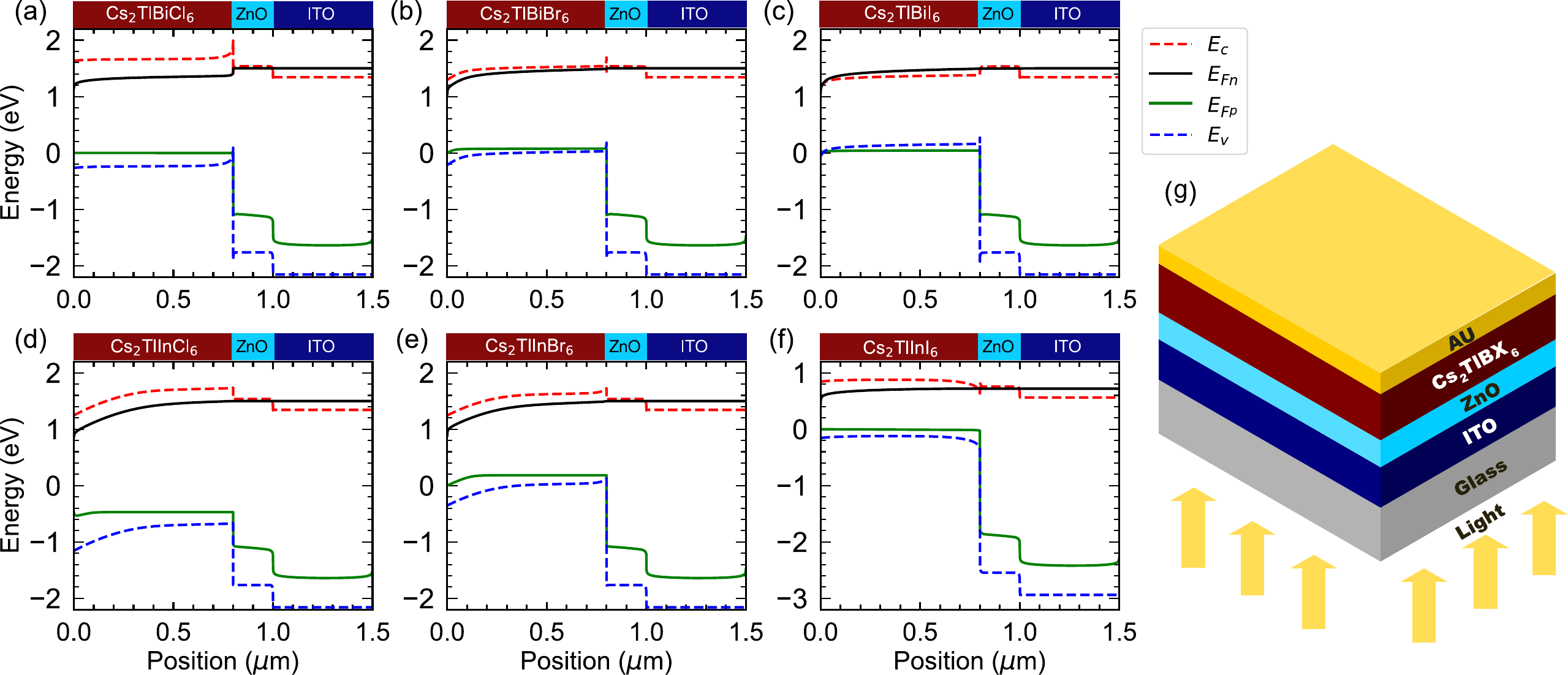}
    \caption{(a-f) Energy band diagram of double perovskites Cs$_2$Tl$BX_6$-based PV simulation and (g) schematic structure of PV device.}
    \label{Fig:EQE}
\end{figure*}

To perform the simulation on SCAPS-1D, we determined the input parameters for all compounds, as listed in Table \ref{tab:param2}. The input parameters for the indium tin oxide (ITO) and ZnO layers were adopted from previous studies, while those for \BX~were derived from our current DFT calculations. For numerical simplicity, the electron and hole thermal velocities in each layer were fixed at $10^7$ cm/s. The electrodes in the front and back were modeled using aluminum (Al) and gold (Au), respectively, as depicted in Fig. \ref{Fig:EQE}(g). The thickness of the perovskite layer plays a critical role in influencing the performance and output parameters of the double-perovskite solar cells because it affects the movement of holes and electrons toward the electrodes.

The energy band diagrams of several \BiX~compounds are presented in Figs. \ref{Fig:EQE}(a--f), showing the minimum and maximum energy levels of the conduction ($E_c$) and valence ($E_v$) bands, respectively. In the device structure, ITO and ZnO serve as the electron transport layer (ETL), while \BiX~acts as the absorber layer. The electron affinity of the ETL must be greater than that of the absorber to facilitate efficient electron transfer from the interface to the absorber \cite{hossain2023designCs}. Since the absorber materials have different bandgaps, their performance is significantly influenced by the choice of ETL. Notably, \BiI~exhibits superior electron and hole transport across the interface layers, as shown in Fig. \ref{Fig:EQE}(c). However, a barrier in the valence band restricts hole transfer from the absorber to the ETL. These interfacial properties play a critical role in enhancing the overall PV efficiency of the device.

\begin{figure*}[ht]
    \centering
    \includegraphics[width=\textwidth]{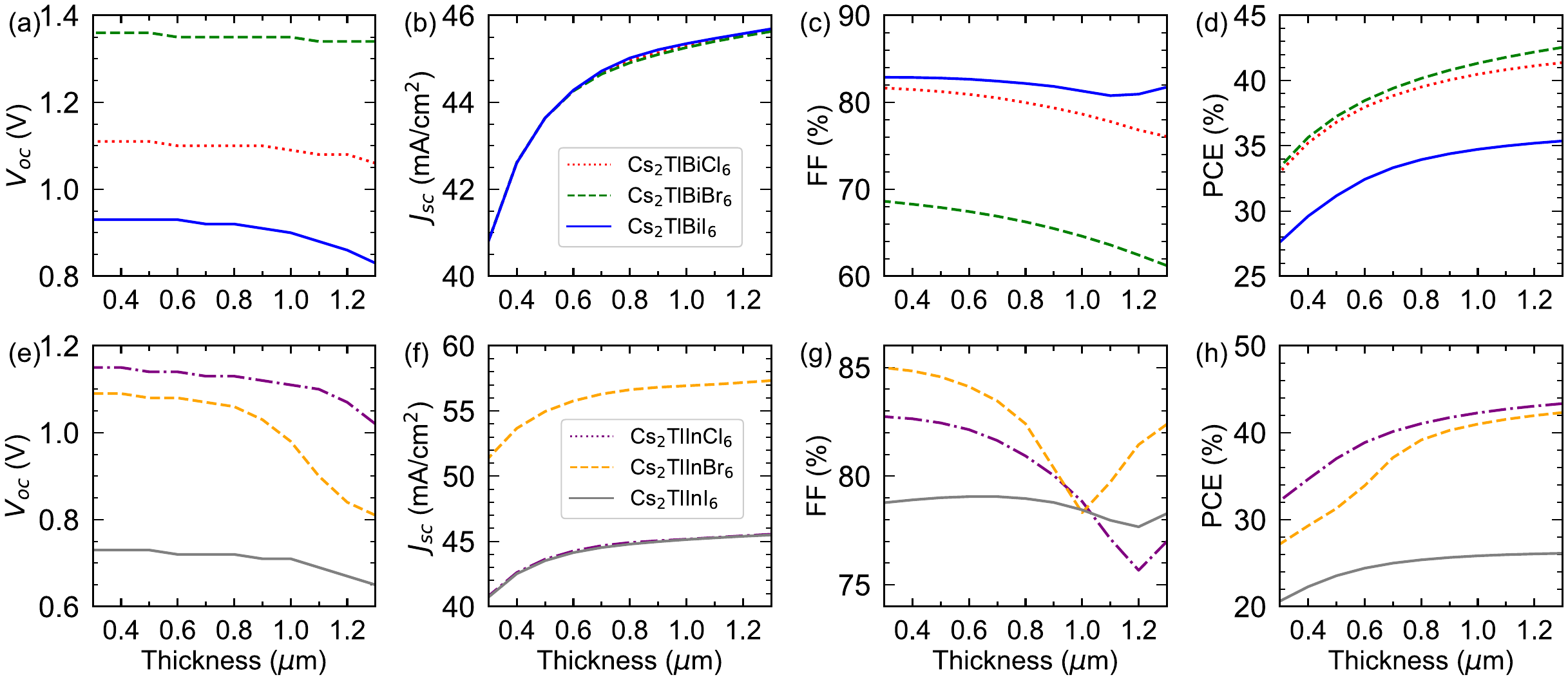}
    \caption{The influence of double perovskites Cs$_2$Tl$BX_6$ thickness on the PV performance parameters: (a,d) open circuit voltage, (b,f) sort-circuit current density, (c,g) fill factor, and (d,h) power conversion efficiency.}
    \label{Fig:thickness}
\end{figure*}

In this study, the thickness of the absorber layer was varied between 0.3 $\mu$m and 1.3 $\mu$m, while all other material parameters were kept constant. A positive correlation was observed between an increase in open-circuit voltage ($V_{\textrm{OC}}$)and the thickness of the \BX~absorber, as shown in Figs. \ref{Fig:thickness}(a) and \ref{Fig:thickness}(e). It suggested that the generation of charge carriers is closely linked to the $V_{\textrm{OC}}$ of solar cells \cite{saha2018boosting}. In an ideal solar cell, $V_{\textrm{OC}}$ occurs when no current flows through the cell, i.e., when the forward bias diffusion current balances the photogenerated current. The maximum $V_{\textrm{OC}}$ is achieved at approximately 1.0 $\mu$m, beyond which a slight decrease is observed. This decline is driven by an increased charge carrier generation in thicker layers, accompanied by a rise in the dark saturation current and enhanced charge carrier recombination rates \cite{sawicka2019simulation}. 

The performance of the short-circuit current density ($J_{\textrm{SC}}$) improves significantly with increasing absorber thickness, as illustrated in Figs. \ref{Fig:thickness}(b) and \ref{Fig:thickness}(f). Thicker absorber layers generate more electron-hole pairs, enhancing the overall efficiency of the device. This explains the observed increase in $J_{\textrm{SC}}$ with absorber thickness. However, beyond a certain thickness, $J_{\textrm{SC}}$ begins to decrease due to an increase in the carrier recombination rate within the absorber layer \cite{patel2021device}. Among the six solar cell structures studied, the \BiI~and \InCl~exhibit the highest $J_{\textrm{SC}}$ values, reaching approximately 45.8 mA/cm$^2$ and 57.2 mA/cm$^2$, respectively.

Figures \ref{Fig:thickness}(c) and \ref{Fig:thickness}(g) illustrate the percentage variation in the fill factor (FF) with changes in absorber thickness. The FF value decreases as the absorber thickness increases, following a consistent trend across all structures. This behavior can be attributed to the increased carrier concentration resulting from greater photon absorption in thicker active layers. Although this enhances the generation of electron-hole pairs and initially improves performance \cite{rai2020analysis}, it also leads to higher recombination rates at larger thicknesses. Meanwhile, the PCE shows a significant increase with the absorber thickness from 2.8\% to 43\% for \BiX~and 20\% to 40\% for \InX, as shown in Figs. \ref{Fig:thickness}(d) and \ref{Fig:thickness}(h). In comparison with the Spectroscopic Limited Maximum Efficiency (SLME) result, the PCE shows higher value than previous report with simple perovskite of \BiX~without ETL \cite{qi2024exploring}. Moreover, the efficiency improves with thickness because thinner layers are unable to absorb sufficient photons. However, beyond 1 $\mu$m, the PCE begins to decline due to increased carrier recombination in thicker films. As a result, 1 $\mu$m is identified as the optimal absorber thickness in this study. 

In order to obtain the maximum possible PCE, the optimization is performed for each parameter. These optimizations include changing the thickness of the ETL/HTL, operating temperature, and carrier mobility. The influence of modifying the thickness of ETL and HTL up to 1 $\mu$m can be seen in Figs. S1 and S2 of the supplementary materials, respectively. It shows that reducing the thickness of the ETL to 0.01 $\mu$m increases the PCE around 1\% for each compound.
There is a slight change in the photovoltaic parameters for all of the HTLs thickness variety. However, the PCE has a slight decrease when the thickness of the HTL is increased. This reduction is due to the fact that the HTL thickness grows larger, which produces the series resistance, and then leads to increases in the recombination rate. 

Figure S3 indicates that optimizing the operating temperature has the most significant effect on performance improvement. After applying temperature, the open circuit voltage decreases and the PCE decreases rapidly about 30\% and 19\% for Cs$_2$TlBi$X_6$ and Cs$_2$TlIn$X_6$, respectively. The temperature influences the mobility and lifetime of charge carriers due to the structural changes. In addition, the optimization of carrier mobility is represented in Figs. S4 and S5, for electron and hole, respectively. Further improvement of $V_{\textrm{OC}}$ and PCE can be obtained when the charge carrier mobility is reduced. 
Modifying the defect density tends to improve the photovoltaic performance \cite{islam2024interface}. Previously, the increase in defect density of CsPbBr$_6$/ZnO interface can reduce the PCE \cite{iqbal2024atomistic}. Figure S6 shows the effect of defect density with a similar result, where the PCE start to decrease after $10^{-15}$ cm$^{-3}$. This defect might reduce the mobility of the electrons from the absorber to ETL. 

\begin{figure*}[ht]
    \centering
    \includegraphics[width=0.8\textwidth]{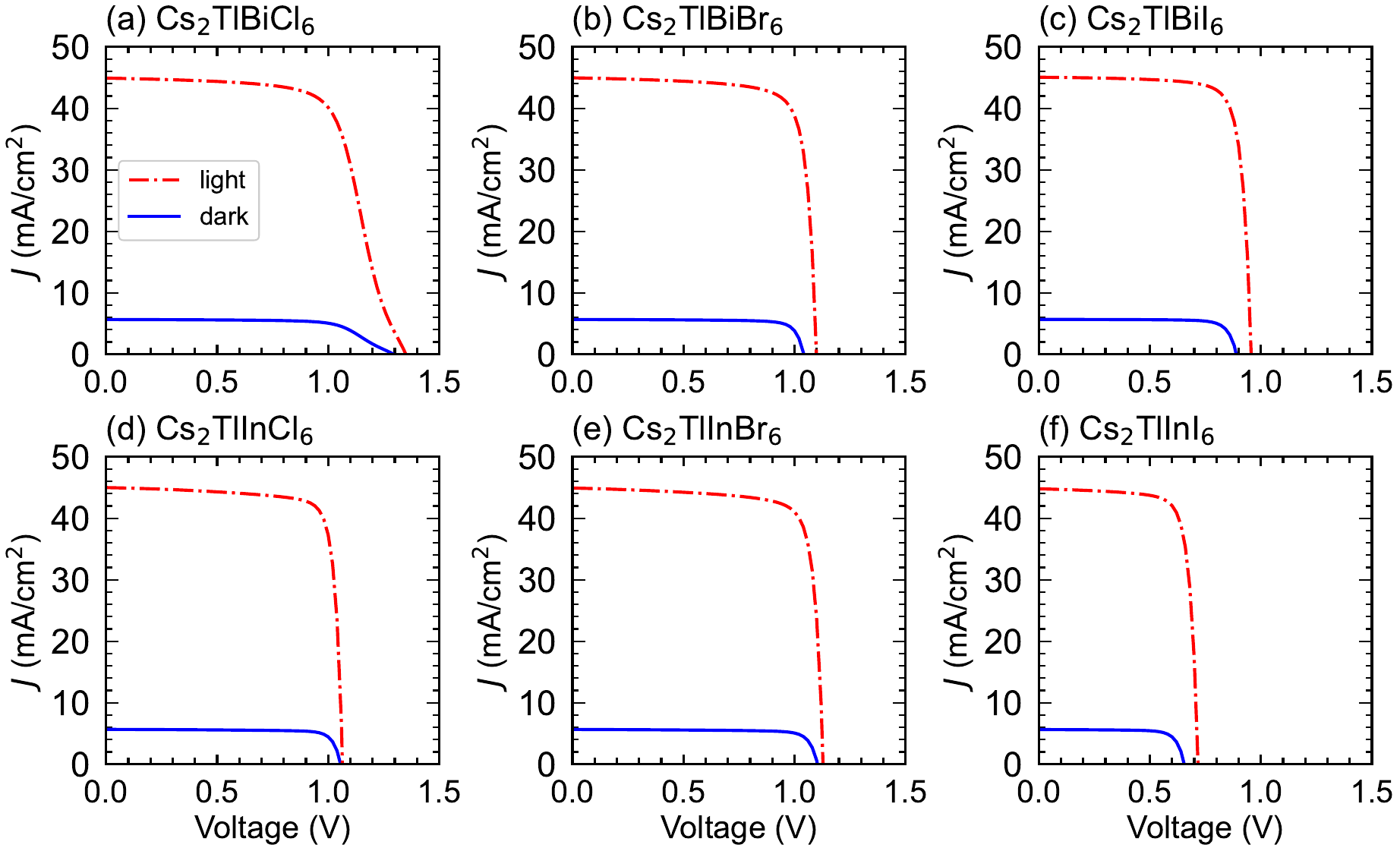}
    \caption{J-V characteristic of some double perovskites \BX.}
    \label{Fig:jv}
\end{figure*}

\begin{figure*}[ht]
    \centering
    \includegraphics[width=0.9\textwidth]{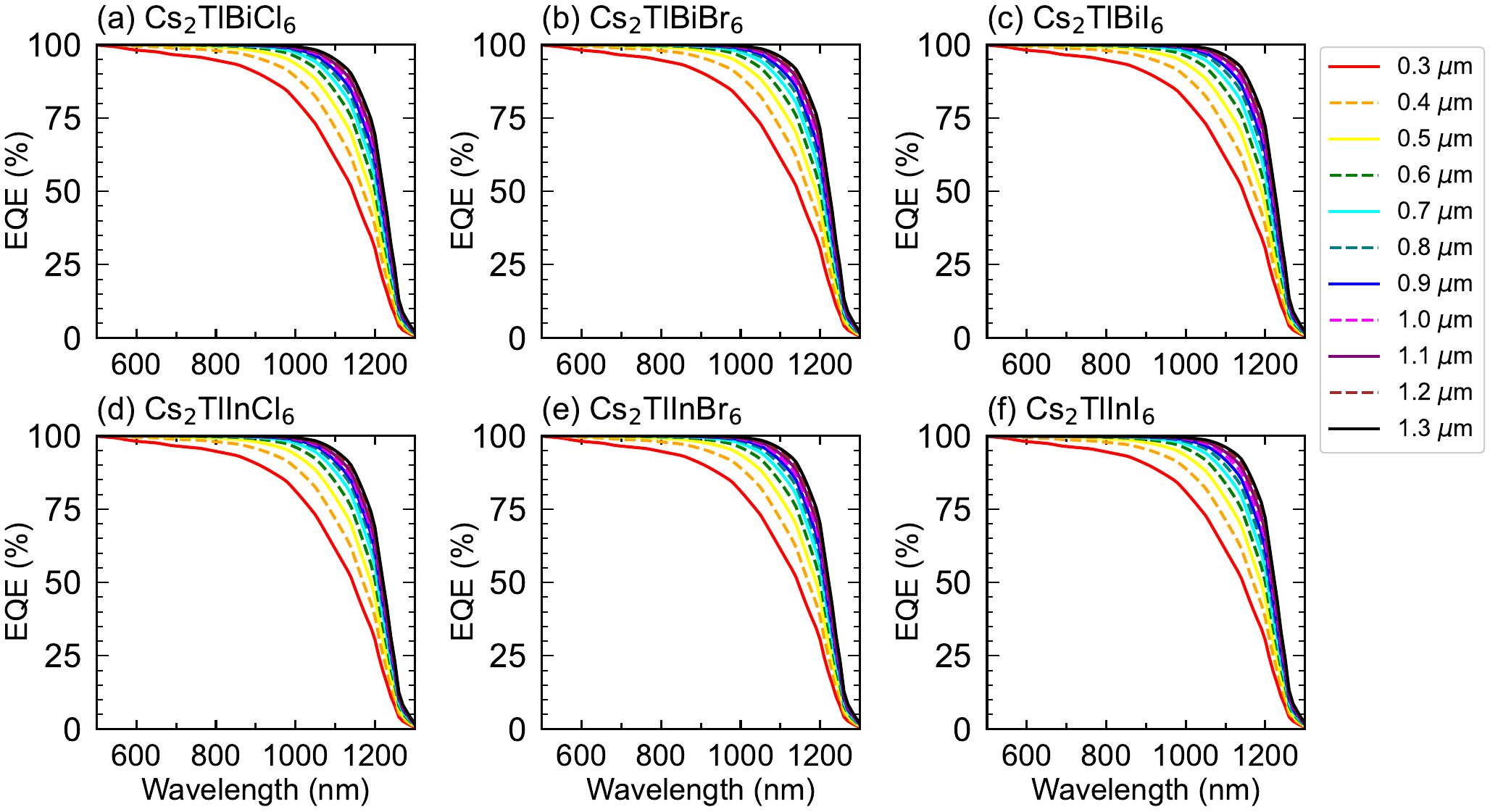}
    \caption{External quantum efficiency (EQE) with various thickness of the absorber layer of \BX.}
    \label{Fig:eqe}
\end{figure*}

The obtained current-voltage characteristics (J-V) of the six optimized double perovskite devices are shown in Fig. \ref{Fig:jv}. A significant decrease is observed in $V_{\textrm{OC}}$ from 1.4 V to 1 V for Cs$_2$TlBi$X_6$ and from 1 V to 0.7 V for Cs$_2$TlIn$X_6$. Taking into account the previous result, the J-V characteristics are slightly affected by the thickness of the layer, leading to an improvement in resistances. Moreover, the external quantum efficiency (EQE) is shown in Fig. \ref{Fig:eqe} where is calculated under the same parameters. On the basis of the analysis, the EQE shifts to a higher wavelength when the thickness is increased. The higher EQE leads to strong absorption and results in enhanced carrier extraction.

\section{Conclusions}
\label{sec:conclution}
We have theoretically investigated the electronic, optical, transport, and PV properties of \ch{Cs_2Tl\textit{BX}_6} double perovskite, where $B=$ Bi, In; and $X=$ Cl, Br, I; through first-principles DFT calculation. Our results confirm that bismuth-based HDPs have direct bandgaps in the range of 1.2--1.9 eV at the $\Gamma$-point, while indium-based HDPs have indirect bandgaps ranging from 2.4--0.8 eV. Furthermore, the bandgap energy decreases with halide substitution from Cl to I, making their bandgaps fall within the energy range of the near-infrared and visible-light spectrum. Among the studied structures, \BiI~exhibits the highest optical absorption in the visible-light range, with a peak value of $5\times10^5$ cm$^{-1}$. In addition, transport analysis reveals that \BiI~has the smallest effective electron mass that leads to high electrical conductivity and high electron mobility, reaching $8\times10^6$ S/m and 120 cm$^2/$V.s, respectively. Further, PV performance analysis shows that \BiX~achieves a promising PCE of up to 42\% at an optimal absorber thickness of 1 $\mu$m. These findings highlight the great potential of \BiX~materials for sustainable solar energy conversion technologies.

\section*{Supplementary data}
Supplementary material related to this article can be found online at 
\url{https://doi.org/10.1016/j.mtcomm.2025.112690}.

\section*{Data availability}
The data required to reproduce these findings are available to download from 
\url{https://hdl.handle.net/20.500.12690/RIN/RVW48A}.

\section*{CRediT authorship contribution statement}
\textbf{Ardimas:} Conceptualization, Formal analysis, Methodology, Investigation, Data Curation, Visualization, Writing--original draft. 
\textbf{E. Suprayoga:} Conceptualization, Formal analysis, Methodology, Validation, Visualization, Writing--review \& editing, Supervision.

\section*{Declaration of Interests}
The authors declare that there are no financial conflicts of interest or personal relationships that could have appeared to influence the work reported in this paper.

\section*{Acknowledgements}
This work was supported by BRIN Talent Management through Postdoctoral Fellowship 2024 Batch 1. The authors acknowledge HPC-BRIN Mahameru for the computational facilities.

\biboptions{sort&compress}

\end{document}